\documentclass[twoside]{article}

\usepackage{arxiv}

\usepackage[utf8]{inputenc} 
\usepackage[T1]{fontenc}    
\usepackage{hyperref}       
\usepackage{url}            
\usepackage{booktabs}       
\usepackage{amsfonts}       
\usepackage{nicefrac}       
\usepackage{microtype}      
\usepackage{lipsum}
\usepackage{fancyhdr}       
\usepackage{graphicx}       

\title{Bias-reduced estimation of mean absolute deviation around the median
}

\author{
  Michele Lambardi di San Miniato \\
  Department of Economics and Statistics \\
  University of Udine \\
  Udine, Italy \\
  \texttt{michele.lambardi@uniud.it} \\
}

\usepackage{subcaption}
\usepackage{natbib}
\usepackage{amsmath}
\renewcommand{\Pr}{\mathbb{P}}
\renewcommand{\check}[1]{\tilde{\tilde{#1}}}

\begin{document}
\maketitle

\begin{abstract}
A bias-reduced estimator is proposed for the mean absolute deviation parameter of a median regression model. A workaround is devised for the lack of smoothness in the sense conventionally required in general bias-reduced estimation. A local asymptotic normality property and a Bahadur--Kiefer representation suffice in proving the validity of the bias correction. The proposal is developed under a classical asymptotic regime but, based on simulations, it seems to work also in high-dimensional settings.
\end{abstract}


\section{Introduction}

The purpose of this note is to present the possibility of bias reduction for Z-estimators even in the absence of some smoothness requirements that are typical in this field \citep[Ch. 5]{Vaart1998}. Bias can be corrected via bootstrap, but this approach can be troubled in high-dimensions \citep{Singh1998}. Here, an interpretable bias-reduced estimator is proposed, in the style of \cite{Firth1993}, for the case of a mean absolute deviation (MAD) parameter of a median regression model, which lacks some conventional smoothness requirements. The proposal is tailored down to MAD around the median, but it can be generalized to other dispersion parameters. In this case, bias reduction involves some unknown model quantities that will be replaced with feasible estimates in the style of \cite{Severini1999}.

The issue of estimation bias in high-dimensions was first pointed out by \cite{Neyman1948}, based on a many-normal-means example. As a response, the classical asymptotic theory, henceforth addressed as first-order asymptotics (FOA), evolved into higher-order asymptotics (HOA). The behavior of square-root-consistent estimators has since been assessed more precisely, leading to bias-reduced estimators \citep{Firth1993}. The style of proof in FOA stems from the work by Cramér and relies heavily on certain local quadratic approximations to the logarithms of likelihood ratios \citep[Sec. 6.1]{Cam2000}. HOA relies on even more detailed stochastic expansions and has thus been built upon stronger smoothness requirements, after the success of Cramér's research line. 

A parallel research line can be found in the work by Le Cam, who aimed at developing a more parsimonious asymptotic theory with weaker assumptions. This theory is based on concepts like contiguity, differentiability in quadratic mean, and local asymptotic normality (LAN), which are less demanding than the smoothness assumptions of HOA \citep[Ch. 6]{Cam2000} \citep[Ch. 6--7]{Vaart1998} \citep{Pollard1997}. Both Cramér's and Le Cam's styles of proof have their merits, as the former is useful at developing \textit{given} some hypotheses, but the latter is needed for \textit{refining} them. Concerns after Le Cam motivate the development of bias reduction in a famous non-smooth estimation problem, namely, quantile regression \citep{Koenker2005}.

Here, we propose an improved estimator of MAD in median regression. A close resemblance with the famous degrees of freedom correction for the variance estimator can be noticed. This correction arises naturally while assessing the LAN property of the criterion that is minimized by the median regression estimator. This LAN condition must be complemented with a Bahadur--Kiefer (BK) representation of the coefficients estimator \citep{Bahadur1966,Kiefer1967}. Both the LAN property and the BK representation are collected in a theorem by Koenker and its proof \citep{Koenker2005}. These two key results, while having been in plain sight for long, have never been combined together to assess and correct the estimator of MAD. Here, we aim to fill in this strange void.

The proposed MAD estimator can be trusted under a classical asymptotic regime, with $p$ fixed, but it is worth testing it in high-dimensional settings too, where both $p$ and $n$ can diverge, in a way such that $n/p\to k \in \,]0,1[$. A grid of values of $p$ and $k$ can be probed, synthetic data can be simulated accordingly and analyzed automatically. Based on simulations, the enhanced estimator seems reliable at least when the true error distribution is known. A distribution-agnostic semi-parametric estimator is also presented, borrowing from empirical corrections \citep{Severini1999}. Under his empirical approach, some model quantities required for bias reduction can be replaced with feasible estimates, though with some \textit{caveats} in high-dimensions. Unfortunately, this replacement looks necessary to correct the MAD estimator in general. The simulation also covers this case and hints at the possibility of an empirical bias reduction without smoothness.

\section{Proposal}

Consider the linear model
$$Y_i = x_i^\top\beta + \gamma \epsilon_i \,,\quad i = 1, \dots, n \,.$$
Here, $\epsilon_i \sim f(\cdot)$ are independent and identically distributed error terms with $\Pr(\epsilon_i \leq 0)=1/2$ and expected absolute value $\mathbb{E}|\epsilon_i|=1$, so that $\mathbb{E}|Y_i-x_i^\top\beta| = \gamma$. All the $x_i$'s are $p$-dimensional covariate vectors, $\beta\in\mathbb{R}^p$ denotes the median regression coefficient vector, $\gamma>0$ is the MAD parameter. MAD is commonly defined as the mean absolute deviation around the \textit{mean}, but its assessment is not as straightforward as it is when defined in terms of \textit{median}. Here the focus is on median regression, for which we define the MAD criterion as
$$\gamma(b) = \frac{1}{n}\sum_{i=1}^n |Y_i - x_i^\top b| \,.$$
The true values of the regression parameter $\beta$ and the MAD parameter $\gamma$ satisfy
$$\beta=\arg\min_b \mathbb{E}\gamma(b)\,,\quad \gamma=\mathbb{E}\gamma(\beta) \,,$$
while the estimator $\hat{\beta}$ for $\beta$ is defined as
$$\hat{\beta}=\arg\min_b\gamma(b)\,.$$

The estimator of $\gamma$ defined as $\bar{\gamma}=\gamma(\beta)$ is unbiased but unavailable in practice, as $\beta$ is unknown, and it is feasibly replaced with $\hat{\gamma}$, defined as $\hat{\gamma}=\gamma(\hat{\beta})$. By design, it holds $\hat{\gamma}\leq \bar{\gamma}$, because $\hat{\gamma}=\min_b\gamma(b)$, so $\hat{\gamma}$ is potentially affected by a downward bias in general. For the sake of asymptotics, assume the absolute value of $\epsilon_i$ has finite variance and set $v=\mathbb{V}|\epsilon_i|$. For instance, it holds $v=\frac{\pi}{2}-1$ and $v=1$ in the case of $\epsilon_i$ distributed as $\mathcal{N}(0,\pi/2)$ and $\mathrm{Laplace}(0,1)$, respectively. Then, it should hold
$$\mathbb{V} \hat{\gamma} \,\dot=\, \mathbb{V} \bar{\gamma} = \frac{v}{n} \,,$$
because $\gamma$ and $\beta$ are orthogonal parameters.

As a matter of notation, let a random sequence $X_n$ be asymptotically of order $o_p(n^\alpha)$ in probability \citep[Section 2.2]{Vaart1998} if the sequence $Y_n=X_n/n^\alpha$ satisfies $\Pr(|Y_n| > \epsilon) \to 0$ in the limit as $n\to\infty$, for all $\epsilon>0$. Expected values of such sequences, if finite, obey the rule $\mathbb{E}o_p(n^\alpha) = o(n^\alpha)$, where $o(n^\alpha)$ is a deterministic sequence $x_n$ such that $\lim_{n\to\infty} x_n/n^\alpha = 0$.

For the main result to hold, it suffices that the proof of Theorem 4.1 by \cite{Koenker2005} can be trusted. Let $X$ be the design matrix, with $x_i^\top$ as its $i$-th row. Then, let $D_0 = X^\top X / n$, $D_1 = D_0 \, f(0)/\gamma$, $W \sim \mathcal{N}(0, D_0/4)$. The Theorem states that the BK representation of $\hat{\beta}$ in median regression is
\begin{equation}\label{eq:bahadur}
	\sqrt{n}(\hat{\beta} - \beta) = D_1^{-1} W + o_p(1) \,.
\end{equation}
which holds in the limit as $n\to\infty$. In the proof of the same Theorem, the function $\gamma(b)$ is shown to satisfy a LAN property. In particular, when setting $b_n=\beta+o_p(n^{-1/2})$, it holds
\begin{equation}\label{eq:lan}
	\frac{n}{2}\left\{\gamma(b_n) - \bar{\gamma}\right\} = -\delta^\top W + \frac{1}{2}\delta^\top D_1 \delta + o_p(1) \,,
\end{equation}
in the limit as $n\to\infty$. Here, $D_1$ serves as a proxy to the Hessian matrix of $\gamma(b)$, which cannot be obtained via direct differentiation. This formulation is compatible with a more general one devised by Boos and Stefanski for the general case of non-smooth estimating functions \citep[Section 7.4]{Boos2013}. Koenker's Theorem with its proof is extremely valuable to our discussion, as it allows to assess the joint distribution of $\hat{\gamma}-\bar{\gamma}$ and $\hat{\beta}-\beta$, based only on shared building blocks, namely, $W$ and $D_1$.

Under Equations \ref{eq:bahadur} and \ref{eq:lan}, the following result holds.
\begin{equation}\label{eq:koenker}
	\hat{\gamma} - \bar{\gamma} = -\frac{1}{n}W^\top D_1^{-1} W + o_p(n^{-1}) \,.
\end{equation}
Let $\chi^2_p$ be a chi-square distributed random variable with $p$ degrees of freedom, such that $\mathbb{E}\chi^2_p=p$ and $\mathbb{V}\chi^2_p=2p$. Equation \ref{eq:koenker} can then be read as
\begin{equation}\label{eq:degfreed}
	\hat{\gamma}-\bar{\gamma} = -\gamma \frac{\chi^2_p / n}{4f(0)} + o_p(n^{-1}) \,,
\end{equation}
which implies that
\begin{equation}\label{eq:leadbias}
	\mathbb{E}\hat{\gamma} = (1 - c)\gamma + o(n^{-1}) \,,\quad\mathrm{with}\quad c= \frac{p/n}{4 f(0)} \,.
\end{equation}
So, when $f(0)$ is known, a bias-reduced estimator $\tilde{\gamma}$ for $\gamma$ can be defined as
$$\tilde{\gamma} = \frac{\hat{\gamma}}{1-c} \,.$$

As a remark, the error distribution $f(\cdot)$ is unknown in general. Any specific assumption in this respect likely leads to a misspecified model. One may replace $f(0)$ with a kernel density estimate (KDE) \citep[Sec. 3.4.1]{Koenker2005}, denoted by $\hat{f}(0)$ and obtained by analyzing the standardized residuals $\hat{\epsilon}_i=(Y_i - x_i^\top\hat{\beta})/\hat{\gamma}$. Under suitable asymptotic guarantees, from semi-parametric statistics, it $\hat{f}(0)=f(0) + o_p(1)$. This replacement is analogous to the empirical method by \cite{Severini1999}. Under the empirical adjustment, the bias-reduced estimator is defined as
$$\check{\gamma} = \frac{\hat{\gamma}}{1-\hat{c}} \,,\quad \text{with} \quad \hat{c}=\frac{p/n}{4\hat{f}(0)} \,.$$
This method is clearly approximate, but it spares some risks of model misspecification and may still work under non-extreme asymptotic regimes.

The proposed bias-reduced estimators for MAD addresses the case of median regression. Here, the main cause of bias is $\hat{\beta}$ minimizing the MAD criterion $\gamma(b)$ more or less aggressively, depending on the balance between $p$ and $n\,f(0)$ in \eqref{eq:leadbias}, which measure model complexity and sample information, respectively. That statement formalizes the minimization pressure on $\hat{\gamma}=\gamma(\hat{\beta})$ that makes it downwardly biased for $\gamma$. This issue is analogous to the loss of degrees of freedom in the estimation of variance parameters, see \eqref{eq:degfreed}. Other estimators for $\beta$ that do not minimize $\gamma
(b)$, like the ordinary least squares estimator, assuming a symmetric error distribution, may lead to a smaller bias in the estimation of $\gamma$: in such a case, the LAN property in Equation \ref{eq:lan} can hold, but the BK representation for $\hat{\beta}$ in Equation \ref{eq:bahadur} must necessarily be revised.

\section{Simulations}

A simulation will be presented, based on normal- or Laplace- distributed responses, with a design matrix that has either normal predictors or dummy variables. The latter design is referred to as analysis of variance (ANOVA). The true parameter values are set as $\beta=0$ and $\gamma=1$ without loss of generality, as the model has scale and location parameters only.

\begin{figure}
	\centering
	\caption{Normal QQ-plots for $Z$ based on 1000 replicates.}
	
	\begin{subfigure}[b]{0.49\textwidth}
		\subcaption{Laplace response, normal design.}
		\label{fig:zstatlaplnorm}
		\includegraphics[width=\linewidth]{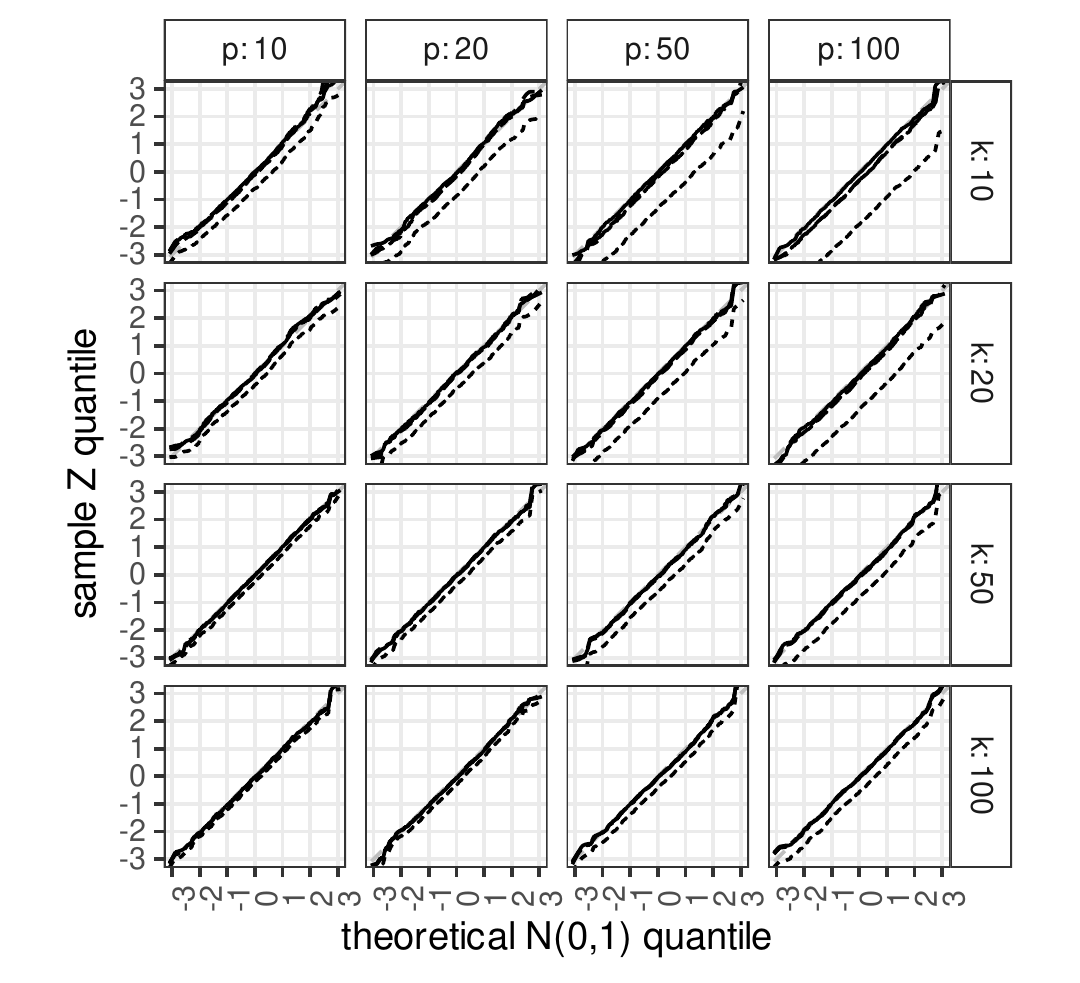}
	\end{subfigure}
	\hfill
	\begin{subfigure}[b]{0.49\textwidth}
		\subcaption{Laplace response, ANOVA design.}
		\label{fig:zstatlaplanova}
		\includegraphics[width=\linewidth]{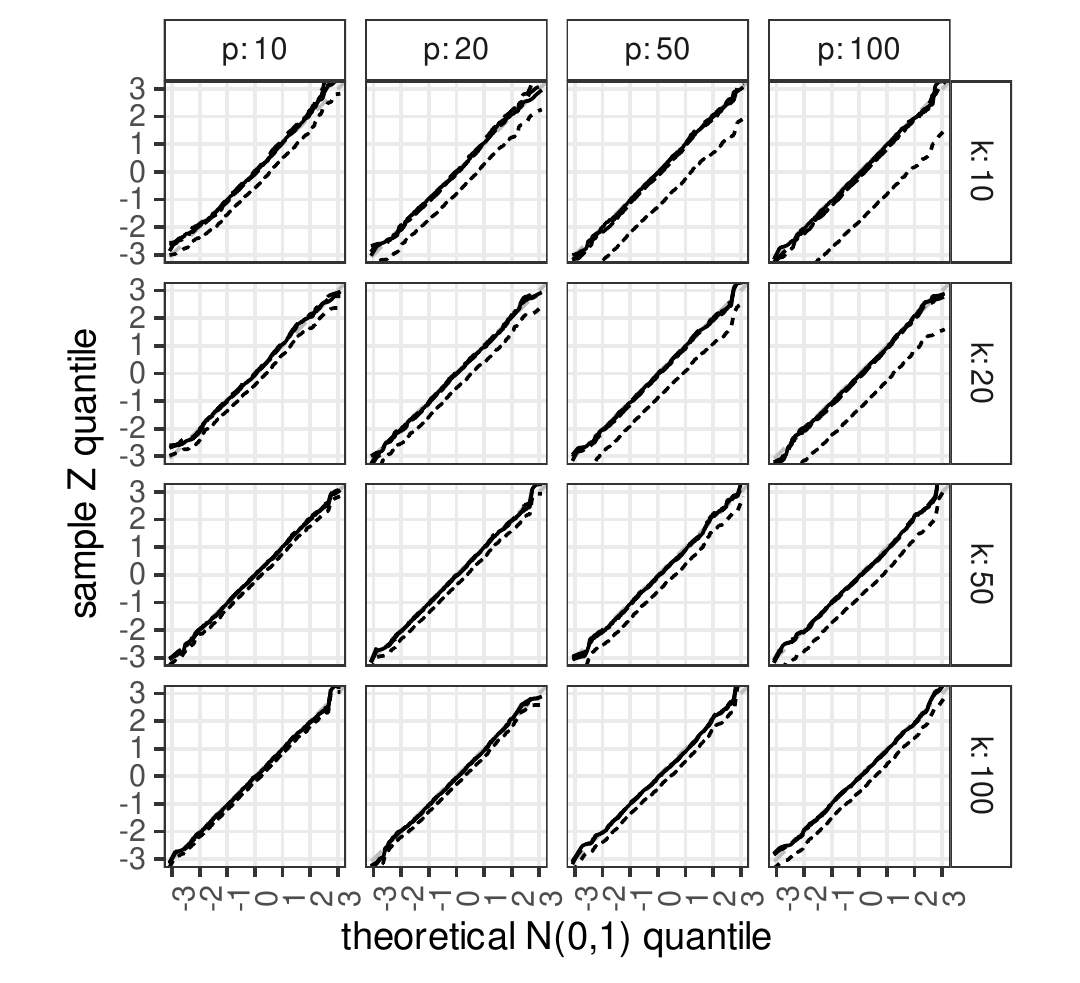}
	\end{subfigure}
	
	\vspace{2em}
	
	\begin{subfigure}[b]{0.49\textwidth}
		\subcaption{Normal response, normal design.}
		\label{fig:zstatnormnorm}
		\includegraphics[width=\linewidth]{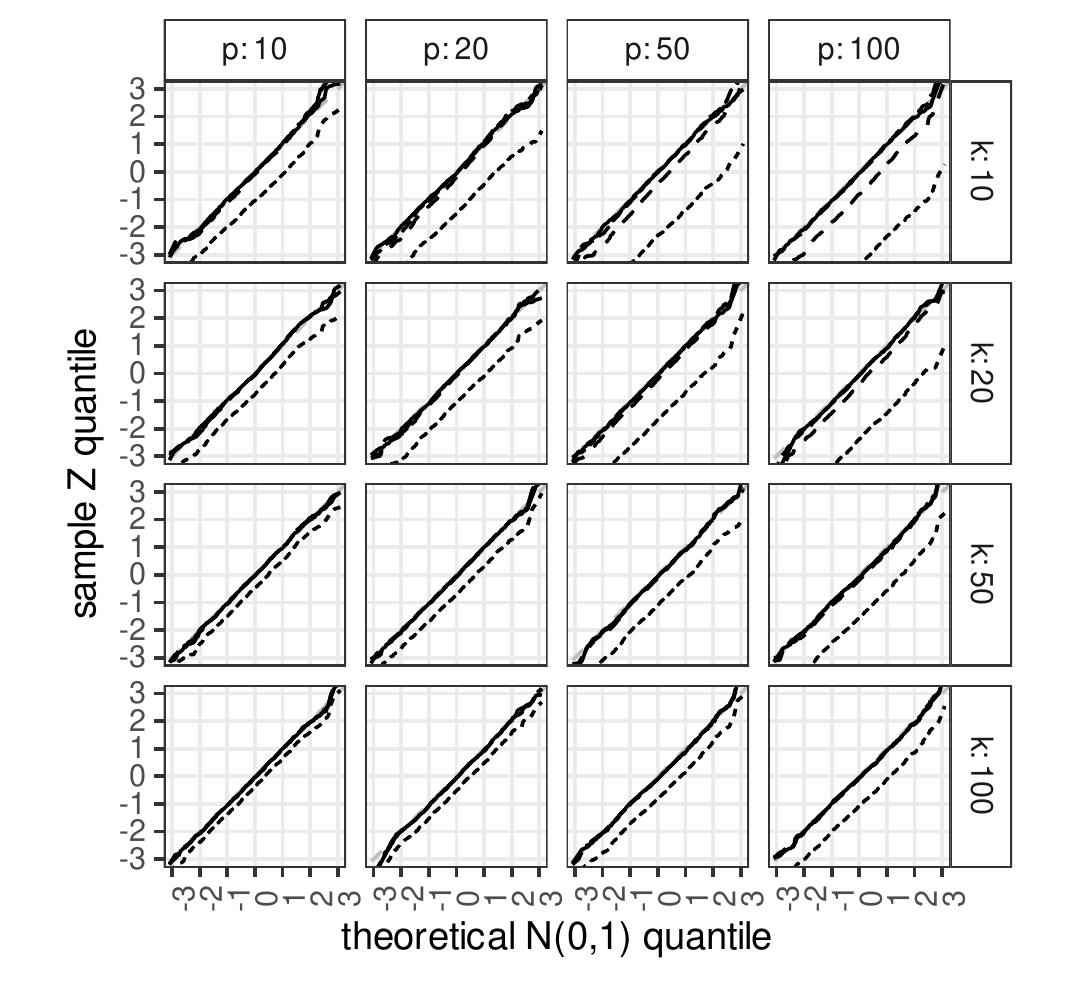}
	\end{subfigure}
	\hfill
	\begin{subfigure}[b]{0.49\textwidth}
		\subcaption{Normal response, ANOVA design.}
		\label{fig:zstatnormanova}
		\includegraphics[width=\linewidth]{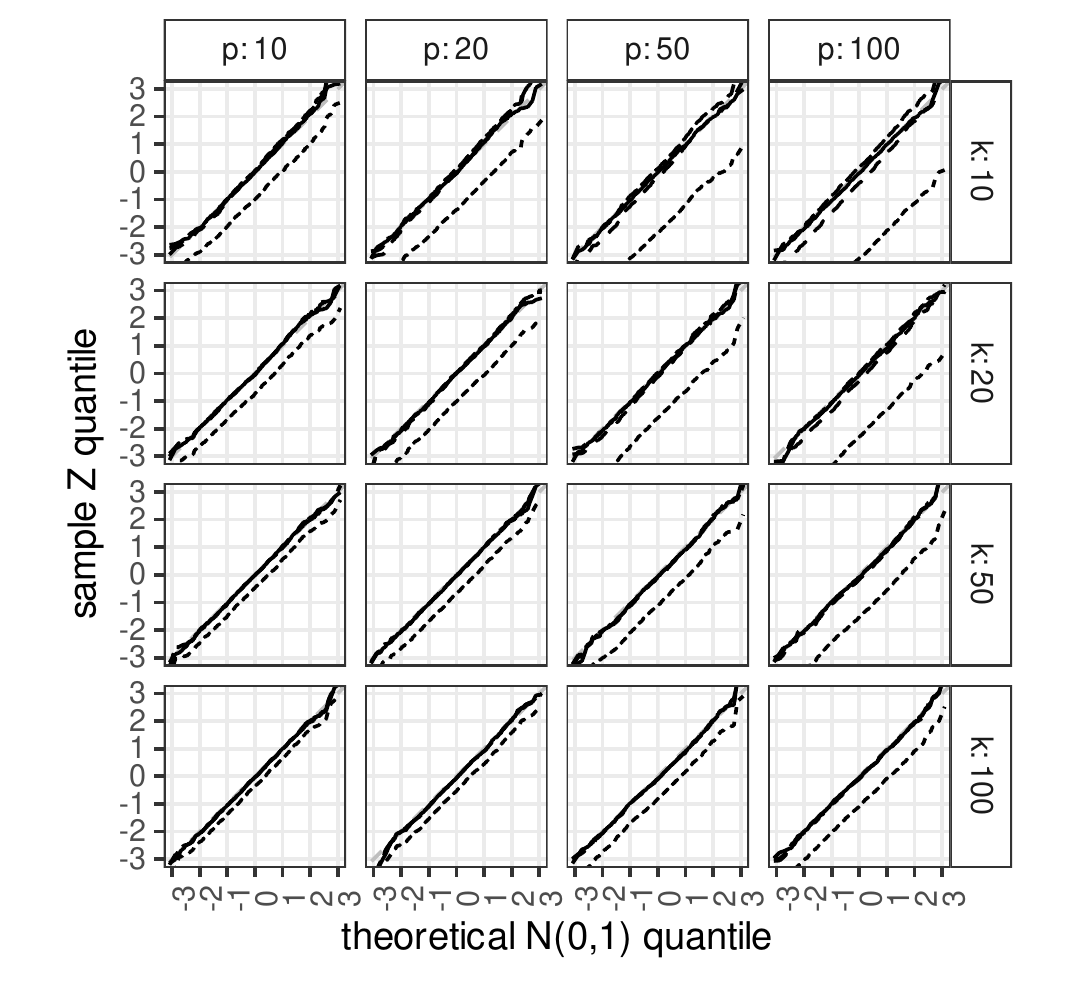}
	\end{subfigure}
	\includegraphics{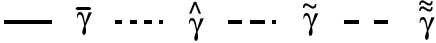}
\end{figure}

In Figures \ref{fig:zstatlaplnorm}-\ref{fig:zstatnormanova}, the normal QQ-plots are reported for a $Z$ statistic defined as
$$Z = \left(\frac{G}{\gamma} - 1\right)/\sqrt{\frac{v}{n}} \, \dot\sim \, \mathcal{N}(0,1) \,.$$
Here, $v$ is the marginal variance of the $\epsilon_i$'s and $G$ is either the unavailable estimator $\bar{\gamma}$, the classical biased estimator $\hat{\gamma}$, the exact bias-reduced estimator $\tilde{\gamma}$ or its empirical counterpart $\check{\gamma}$. Each estimator is represented by a curve with a different linetype. The curves ideally lie close to the dashed grey diagonal of each panel, but otherwise it is desirable that they lie close to the curve for the ideal estimator $\bar{\gamma}$.

The most relevant aspects in the simulation are clearly the number of regression parameters $p$ (on the $x$-axis) and the sample information ratio $k=n/p$ (on the $y$-axis). The bias-reduced estimators were derived and assessed only under the classical asymptotic regime, which is mimicked by reading a column of plots from top to bottom. Moderate $p/n$ regimes can be approximately observed within the rows of plots, by reading from left (small $p$ and $n$) to the right (large $p$ and $n$). The proposed bias-reduced estimators looks more reliable under the classical regime, while the moderate $p/n$ regimes are visibly more challenging, but they still perform quite well nonetheless.

The $Z$ statistic for both the enhanced estimators $\tilde{\gamma}$ and $\check{\gamma}$ looks approximately $\mathcal{N}(0,1)$. The empirical bias-reduced estimator $\check{\gamma}$ catches up with the performance of the exact estimator $\tilde{\gamma}$. This result suggests that the empirical adjustment is preferable, since more robust to model misspecification. The main drawback is that $f(0)$ is estimated, so $\check{\gamma}$ provides a less thorough bias reduction than allowed by $\tilde{\gamma}$ when the true error distribution is known.

Estimation bias seems unaffected by the design or by the error distribution, at least as it concerns the enhanced estimators $\tilde{\gamma}$ and $\check{\gamma}$. On the contrary, the magnitude of bias in the case of $\hat{\gamma}$ depends more heavily on the error distribution, in line with \eqref{eq:leadbias}. In the most extreme situations, with small $k$ and large $p$, also the empirical correction $\check{\gamma}$ may be questioned, depending on design and error distribution, but it is still preferable over the unadjusted estimator $\hat{\gamma}$.

\section{Discussion}

In this note, the focus was on the MAD parameter $\gamma$ in the median regression case, where $\Pr(Y_i \leq x_i^\top \beta) = 1/2$ and $\mathbb{E}|Y_i-x_i^\top\beta|=\gamma$. The median regression case is especially interesting when assessing inference on the MAD parameter, because the related MAD criterion is subject to a minimization pressure that makes it systematically upper bounded by the unbiased estimator $\bar{\gamma}$. The assessment is different, but related, when considering MAD around the mean, since the ordinary least squares estimator has a distinct (and exact) BK representation in place of \eqref{eq:bahadur}. The LAN property in \eqref{eq:lan} is preserved if $x_i^\top\beta$ coincides with the median of $Y_i$, for instance, when the error distribution is symmetric.

An adaptation of Severini's empirical bias-reduced estimation was illustrated, for the case when some model quantities required for the enhancement, like $f(0)$, are unknown. A cautionary tale is in order, since the exact adjustment $\tilde{\gamma}$ and the empirical counterpart $\check{\gamma}$ are not perfectly exchangeable, as also seen in simulation. Differences in performance are visible especially under moderate $p/n$ regimes. There is thus an open issue in this research, concerning the way to mimic the stronger bias reduction attained by the exact adjustment $\tilde{\gamma}$, but the empirical correction $\check{\gamma}$ may still be trusted under less extreme asymptotic regimes.

\end{document}